\documentclass[12pt]{amsart}
\usepackage{amsthm} 
\usepackage{amsfonts}
\begin{document} 
\title{Quadratic Bosonic and Free White Noises} 
\author{Piotr \'Sniady}
\thanks{Research partially supported by the Scientific Research 
Committee in Warsaw under grant number P03A05415.} 
\address{Instytut Matematyczny, Uniwersytet Wroc\l{}awski, 
pl.~Grunwaldzki 2/4,\\ 
50-384 Wroclaw, Poland}   
\email{psnia@math.uni.wroc.pl}
\date{Received: 23 August 1999 / Accepted: 8 December 1999}
\maketitle 
\begin{abstract}
We discuss the meaning of renormalization used for deriving qua\-dra\-tic 
bosonic commutation relations introduced by Accardi \cite{ALV} 
and find a representation of these 
relations on an interacting Fock space. 
Also, we investigate classical stochastic processes which can be
constructed from noncommutative quadratic white noise. We postulate 
quadratic free white noise commutation relations and find their
representation on an interacting Fock space. 
\end{abstract} 

\newcommand{\Ha}{{\mathcal H}}
\newcommand{\gwia}{^{\star}}
\newcommand{\A}{{\mathcal A}}
\newcommand{\C}{{\mathbb C}}
\newcommand{\R}{{\mathbb R}}
\newcommand{\N}{{\mathbb N}}
\newcommand{\bs}{{\tilde{b}}}
\newcommand{\Gammaq}{{\Gamma^2}}
\newcommand{\Gammaqb}{{\Gamma^2_{\rm b}}}
\newcommand{\Gammaqf}{{\Gamma^2_{\rm f}}}
\newcommand{\Ldwa}{{{\mathcal L}^2}}
\newcommand{\Linf}{{{\mathcal L}^{\infty}}}
\newcommand{\stimes}{\widehat{\otimes}}

\theoremstyle{plain}
\newtheorem{lemma}{Lemma}
\newtheorem{theorem}[lemma]{Theorem}
\newtheorem{proposition}[lemma]{Proposition}
\newtheorem{corollary}[lemma]{Corollary}
\newtheorem{conjecture}[lemma]{Conjecture}
\newtheorem{claim}[lemma]{Claim}
\newtheorem{fact}[lemma]{Fact}

\theoremstyle{definition}
\newtheorem*{definition}{Definition}

\theoremstyle{remark}
\newtheorem*{remark}{Remark}
\newtheorem*{example}{Example}
\newtheorem*{question}{Question}

\section{Introduction}
Hudson and Parthasarathy \cite{HuP} showed that a Brownian motion $B(T)$ can 
be represented as a sum of two noncommuting operators: annihilation 
$a_{(0,T)}$ and creation $a\gwia_{(0,T)}$,
$$B(T)=a_{(0,T)}+a\gwia_{(0,T)}=\int_0^T (a_t+a\gwia_t),$$
where $a_t$ and $a\gwia_t$ stand for the 
infinitesimal annihilation and creation operators respectively. 

Accardi \cite{ALV}, in order to study some physical problems, introduced 
qua\-dra\-tic white noise operators, which informally can be written as 
$n_t=a\gwia_t a_t$, $b_t=(a_t)^2$ and $b\gwia_t=(a\gwia_t)^2$. The first 
one, called the number operator has been already considered in the white 
noise calculus and it does not cause serious difficulties. The 
other two, called quadratic annihilation and quadratic creation 
operators, in fact represent infinite quantities and therefore have to be 
redefined. Indeed, it can be shown that because of 
$[a_t,a\gwia_s]=\delta(t-s)$ we have $$[a_t^2, a\gwia_s{}^2]=2 
\delta^2(t-s)+4 \delta(t-s) a_t a\gwia_s,$$ where $\delta$ denotes the Dirac 
distribution. Since the square of the delta function is not well defined, 
this relation is meaningless. Furthermore it is too singular to apply
the subtraction renormalization \cite{S}.
By renormalization $\delta^2(x)=\gamma_0 
\delta(x)$ Accardi postulates that the renormalized quadratic white noise 
operators should fulfill the following commutation relation: 
\begin{equation}
[b_t, b\gwia_s]=2 \gamma_0 \delta(t-s)+4 \delta(t-s) n_s,
\label{eq:wstep} \end{equation}
which  for smeared operators $b_\phi=\int \overline{\phi_t} b_t$, $b\gwia_\psi
=\int \psi_s b_s$ takes the form
\begin{equation} [b_\phi, b\gwia_\psi] 
 =2 \gamma_0 \langle \phi,\psi \rangle +4n_{\bar{\phi} \psi}.
\label{eq:wstep1} \end{equation}

In Sect.\ \ref{sec:renormalizacja} we present another discussion 
of this relation in a more general context of $q$-deformed commutation 
relations.  

In Sect.\ \ref{sec:bosonic} we show that from this 
discussion follows for the bosonic 
case the meaning of renormalization constant 
$\gamma_0$ as the inverse of the lengthscale taken for quadratic variation 
of a (noncommutative) Brownian motion and we discuss other commutation 
relations. Furthermore, from quadratic white noise operators we construct 
some classical stochastic processes.

Accardi and Skeide \cite{ALV,AS} have constructed a Fock representation
of quadratic white noise relations. The construction presented in the paper 
\cite{ALV} uses the Kolmogorov decomposition for a certain positive kernel. 
Another approach is presented 
in the paper \cite{AS} where the construction of quadratic white noise 
operators is based on the theory of Hilbert modules. 
In Sect.\ \ref{sec:realization} and \ref{sec:another} 
we present a direct construction of 
such a representation on an interacting 
Fock space. Our method is based on defining explicitly a scalar 
product on a symmetric Fock space. 

In Sect.\ \ref{sec:quadraticandlinear} we discuss the existence of a 
Fock representation of an algebra containing both quadratic and usual linear 
white noise operators. It turns out that it is in general not possible to 
find such a representation. The main reason is that under a certain lengthscale 
the renormalized quadratic operators lose their intuitive meaning as squares of
creation and annihilation operators.

In Sect.\ \ref{sec:free} we introduce free quadratic white noise 
operators which should describe the squares of free creation and annihilation 
operators with small violation of freeness and construct their representation.

Both standard and quadratic white noises are weak processes, i.e.\ mappings
from some linear space $S$ to operators on a Hilbert space.
Contrary to white noise commutation relations, the quadratic 
relation (\ref{eq:wstep1}) involves not only a scalar product in $S$,
but a product of two elements of $S$ as well. From the noncommutative
geometry viewpoint \cite{C} 
it would be interesting to consider noncommutative
spacetime algebras $S$ as well and quadratic white noise relations provide
appropriate examples. Unfortunately, for the bosonic white noise there seems
to be some limitations on the choice of $S$ but for the free case the
construction works for all associative algebras.

\section{General Renormalized Quadratic White Noise}
\label{sec:renormalizacja}
For a Hilbert space $\Ha$ and a real number $q$, $-1<q\leq 1$ let us 
consider $q$-deformed white noise operators \cite{FB,BKS}: the creation 
$a_\phi\gwia$ and its adjoint annihilation $a_\phi$ indexed by $\phi\in\Ha$.
These operators fulfill the following commutation relation:
\begin{equation} a_{\phi} a\gwia_{\psi}-q a\gwia_{\psi} 
a_{\phi}=\langle \phi,\psi \rangle. \label{eq:r1} \end{equation}
%

For the case $\Ha={\mathcal L}^2(M,d\mu)$ we can write 
informally 
$$a_\phi=\int_M \overline{\phi(t)} a_t, \qquad a\gwia_\phi=\int_M \phi(t) 
a\gwia_t,  $$ where $a_t$, $a\gwia_t$ denote white noise annihilation and 
creation operators. 

Our goal is to introduce operators $b_\phi$ and $b_\phi\gwia$ which would be 
informally treated as integrals of squares of white noise operators
 $$b_\phi=\int_M \overline{\phi(t)} (a_t)^2, 
\qquad b\gwia_\phi=\int_M \phi(t) (a\gwia_t)^2.  $$
In order to give meaning to these expressions
let us consider a sequence $(I_i)$ of disjoint measurable subsets of 
$M$, each of the same measure $l$ and a sequence $(\chi_i)$ of orthogonal 
functions
 $$\chi_i(x)=\left\{  \begin{array}{ccl} 1 & : & x\in I_i \\ 0 & : & 
x\notin I_i \end{array} \right. .$$ 
Furthermore let us consider piecewise constant 
functions $\phi$, $\psi$, 
$$\phi(x)=\sum_i \phi(x_i) \chi_i(x), \qquad 
\psi(x)=\sum_i \psi(x_i) \chi_i(x), $$ 
for a sequence $(x_i)$ such that $x_i\in I_i$. 
Now let us define 
$$b_\phi=\sum_i \overline{\phi(x_i)} (a_{\frac{1}{\sqrt{l}} \chi_i})^2, 
\qquad b\gwia_\psi=\sum_i \psi(x_i) (a\gwia_{\frac{1}{\sqrt{l}} \chi_i})^2. 
$$ 

A simple computation shows that for squares of creation and annihilation 
operators
$$ a_{\zeta}^2 a\gwia_{\xi}{}^2-q^4 a\gwia_{\xi}{}^2 a_{\zeta}^2=
(1+q) \langle \zeta, \xi \rangle^2 + q (1+q)^2 
\langle\zeta,\xi \rangle a\gwia_{\xi} a_{\zeta} $$
hold. For this reason we have 
$$  b_\phi b\gwia_\psi-q^4 b\gwia_\psi 
b_\phi= $$ $$ =(1+q) \sum_i \psi(x_i) \overline{\phi(x_i)}+ 
q (1+q)^2 \sum_i \psi(x_i) \overline{\phi(x_i)}\  a_{\frac{1}{\sqrt{l}} 
\chi_i}\gwia\ a_{\frac{1}{\sqrt{l}} \chi_i}. $$ 
Since the ${\mathcal L}^2(M,d\mu)$ norm of the function $\frac{1}{\sqrt{l}} 
\chi_i$ is equal to $1$, the operator $a_{\frac{1}{\sqrt{l}} 
\chi_i}\gwia a_{\frac{1}{\sqrt{l}} \chi_i}$ is a number 
operator. If we consider only the creation and annihilation operators 
$a\gwia_\theta$, $a_\theta$ for functions $\theta$ that are piecewise 
constant on sets $I_i$ then the operators $a_{\frac{1}{\sqrt{l}} 
\chi_i}\gwia a_{\frac{1}{\sqrt{l}} \chi_i}$ and $\int_{I_i} a\gwia_t a_t$ 
have the same commutation relations with others and therefore they are 
indistinguishable in the sense of vacuum expectation values. Under these
assumptions we can write \begin{equation}  \label{eq:sss} b_\phi 
b\gwia_\psi-q^4 b\gwia_\psi b_\phi= \end{equation} 
$$= \frac{1+q}{l} \int_M \psi(x) 
\overline{\phi(x)}\ d\mu(x)+q (1+q)^2 \int_M \psi(x) \overline{\phi(x)}\ 
a\gwia_t a_t.$$

The preceding calculations hold only for a very limited class of functions 
$\phi$ and $\psi$. However, we shall postulate the following commutation 
relation between quadratic creation and annihilation operators for all 
$\phi$ and $\psi$:
 \begin{equation} b_\phi b\gwia_\psi-q^4 b\gwia_\psi 
 b_\phi 
 =\gamma \langle \phi,\psi \rangle +c\ n_{\bar{\phi} \psi}, \label{eq:t0} 
\end{equation}
for some constants $\gamma$, $c$ and where $n_f$, called a number 
operator, should be understood as a generalization of the usual 
number operator $\int_M f(x) a\gwia_t a_t$. 


\subsection{Fock representations}
Hudson--Parthasarathy's operators $a_\phi$, $a\gwia_\phi$ ($\phi\in\Ha$) are 
usually represented as operators acting on some Hilbert space with a cyclic 
vector $\Omega$ with the property that $a_\phi \Omega=0$ 
for all $\phi\in\Ha$. 
Since the operators $b_\phi$, $b\gwia_\phi$ are interpreted as smeared
renormalised squares of white noise operators $a_t$, $a\gwia_t$, therefore 
it is natural to ask if it is possible to find a representation
of operators $b_{\phi}$, $b\gwia_{\phi}$, $n_{\phi}$ acting on some Hilbert 
space $\Gammaq$ such that $\Gammaq$ contains a cyclic vector $\Omega$, called 
a vacuum, such that $b_\phi\Omega=0$, $n_\phi\Omega=0$ for all $\phi$. 
In such a setup we will be able to define a state $\tau$ on the space of
operators acting on $\Gammaq$ defined by $\tau(X)=\langle 
\Omega, X \Omega\rangle$ which would play the role of a (noncommutative) 
expectation value.

%
%

\section{Bosonic Quadratic White Noise}
\label{sec:bosonic}
\subsection{Bosonic commutation relations}
For the bosonic case $q=1$ Eq.\ (\ref{eq:sss}) takes the form
\begin{equation} [b_\phi, b\gwia_\psi] 
 =2 \gamma_0 \langle \phi,\psi \rangle +4n_{\bar{\phi} \psi},
\label{eq:t0prim} \end{equation}
where $\gamma_0=\frac{1}{l}$. 
Furthermore, we postulate that two creation, two annihilation and two 
number operators should commute:
\begin{equation} [b_\phi,b_\psi]=0,\qquad [b_\phi\gwia,b_\psi\gwia]=0,\qquad 
[n_\phi,n_\psi]=0. 
\label{eq:t1}
\end{equation}

A simple calculation for piecewise constant functions 
$$\biggl[ \sum_i \phi(x_i) a\gwia_{\frac{1}{\sqrt{l}} \chi_i} 
a_{\frac{1}{\sqrt{l}} \chi_i},
\sum_j \psi(x_j) (a\gwia_{\frac{1}{\sqrt{l}} \chi_j})^2 \biggr]=2 
\sum_k \psi(x_k) \phi(x_k) (a\gwia_{\frac{1}{\sqrt{l}} \chi_k})^2 $$ 
gives us a motivation for the following commutation relations: 
\begin{equation} [n_\phi, b\gwia_\psi]=2 
b\gwia_{\phi\psi},\qquad [b_\psi, n_\phi]=2 b_{\psi 
\phi\gwia}. \label{eq:t2prim} \end{equation}

\subsection{Classical quadratic processes}
By the spectral theorem a commuting family of normal operators has a common 
spectral measure. After applying a state the spectral measure becomes an 
ordinary measure which has a natural probabilistic interpretation as a joint 
distribution of random variables corresponding to operators from our family.

Let us define for $s\in\R$, 
\begin{equation} Q_s(\phi)=b_{\phi\gwia}+b\gwia_\phi+s n_{\phi}.
\label{eq:qs} \end{equation}
Similar to the white noise it is a weak process \cite{S}, i.e.\ an 
operator valued function on a linear space $\Ldwa(M,d\mu)\cap \Linf(M,d\mu)$.
In the case $M=\R_+$ we can construct from it a stochastic process
$Q_s(t)=Q_s(\chi_{(0,t)} )$.
\begin{theorem}
Let us fix $s\in\R$. Then $\{Q_s(\phi)\}$ 
forms a commuting family of normal operators and therefore it is a 
classical stochastic process. With respect to the expectation value $\tau$, 
it is a Markovian process.
\end{theorem}
\begin{proof}
The first part of the proof is a simple application of 
(\ref{eq:t0prim})--(\ref{eq:t2prim}). 

The property that $Q_s$ is a process with independent increments
means exactly that for all disjoint sets $M_1, 
M_2\subset M$ and $f_i\in\mbox{Alg}\{Q_s(\phi):\phi\in 
{\mathcal L}^2(M_i)\}$ the equality
$\tau(f_1 f_2)=\tau(f_1) \tau(f_2)$ holds.
Note that every expression containing operators $n_\phi$, $b_\phi$, 
$b\gwia_\phi$ ($\phi\in\Ha$) can be written according to the relations 
(\ref{eq:t0prim}), (\ref{eq:t2prim}) in the normal form, a linear 
combination of products of type
\begin{equation} \label{eq:normal} b\gwia_{\phi_1} \cdots b\gwia_{\phi_k} 
n_{\chi_1} \dots n_{\chi_m} b_{\psi_1}\cdots b_{\psi_l} . \end{equation}
Each of the operators $n_{\phi_1}$, $b_{\phi_1}$, $b\gwia_{\phi_1}$ 
commutes with each of the operators  $n_{\phi_2}$, $b_{\phi_2}$, 
$b\gwia_{\phi_2}$ for $\phi_i\in {\mathcal L}^2(M_i,d\mu)$, therefore 
a product of two expressions of the form 
(\ref{eq:normal}), one being an element of $\mbox{Alg}\{n_\phi, 
b_\phi, b\gwia_\phi:\phi\in {\mathcal L}^2(M_1)\}$ and the other an element 
of $\mbox{Alg}\{n_\phi, b_\phi, b\gwia_\phi:\phi\in {\mathcal L}^2(M_2)\}$ 
is--up to a permutation of factors--in a normally ordered form.  
The state $\tau$ has a property that on normally ordered products it 
takes nonzero values only on multiples of identity and 
$\tau(f_1 f_2)=\tau(f_1) \tau(f_2)$ follows.

Now it is enough to notice that the expectation value of $Q_s(\phi)$ is
equal to $0$ for any $\phi$.
\qed \end{proof}

\subsection{Quadratic variation of a Brownian motion}
\label{subsec:variation}
Let $M=\R_+$ and let us consider an arithmetic series $(t_i)$, $t_i=l i$. 
For the sum of squares of increments of a standard Brownian motion the 
following operator equality holds:
 $$\sum_i \phi(t_i) [B(t_{i+1})-B(t_i)]^2=\sum_i 
\phi(t_i) [a_{\chi_{i}}+a\gwia_{\chi_{i}}]^2=$$ 
$$= \sum_i \phi(t_i) [a_{\chi_{i}}^2+2 a_{\chi_{i}}\gwia a_{\chi_{i}}+ 
a_{\chi_{i}}\gwia{}^2+(t_{i+1}-t_i)],$$ 
where $\chi_i$ is the characteristic function of an interval 
$(t_i,t_{i+1})$. In the preceding discussion we have chosen the commutation 
relations between operators $l b_{\chi_i}$, $l b\gwia_{\chi_j}$ and $l 
n_{\chi_k}$ to coincide with commutation relations between $a_{\chi_i}^2$, 
$a^{\star 2}_{\chi_j}$ and $a\gwia_{\chi_k} a_{\chi_k}$ whenever the length of 
intervals is equal to $l=\frac{1}{\gamma_0}$.
Therefore, for any function $\phi$ which is piecewise constant on intervals 
$(t_i,t_{i+1})$ we can write $$\sum_i \phi(t_i) [B(t_{i+1})-B(t_i)]^2=$$ 
$$=\sum_i \phi(t_i) \left\{ \frac{1}{l} 
[b_{\chi_{i}}+b\gwia_{\chi_{i}}+2n_{\chi_{i}}] +(t_{i+1}-t_i) 
\right\}= \frac{1}{l} Q_2(\phi)+\int_{\R_+} \phi(x)\ dx. $$
This equation can be viewed as follows. Just like $a_\phi$, $a\gwia_\phi$ 
are quantum components of the Brownian motion, for functions $\phi$ which 
are piecewise constant on intervals which length is a multiplicity of 
$\frac{1}{\gamma_0}$ operators $b_\phi$, $b\gwia_\phi$, $2n_\phi$ are quantum 
components of the quadratic variation of Brownian motion. 
The constant $\frac{1}{\gamma_0}$ describes the lengthscale under which 
such interpretation is no longer valid.

The measures corresponding to $\gamma_0 Q_2(t)+t=
\gamma_0 Q_2(\chi_{(0,t)})+t$ for $t$ being the 
multiplicity of $\frac{1}{\gamma_0}$ are therefore the $\chi^2$ distributions.
From this it follows that for arbitrary $t$ these are gamma distributions
and $\gamma_0 Q_2(t)+t$ is a gamma process.


\section{Quadratic Bosonic White Noise on an Interacting Fock Space}
\label{sec:realization}
Let $\A$ be a commutative $C\gwia$-algebra of continuous functions on some
set $M$ with a measure $\mu$ and let the state on $\A$ 
induced by $\mu$ be denoted also by $\mu$.

\begin{definition}
A partition of a finite set $A$ is a collection
$\pi=\{\pi_1,\dots,\pi_m\}$ of nonempty sets $\pi_p$, which are pairwise
disjoint and their union is equal to $A$.

An ordered partition of a finite set $A$ is a set 
$\pi=\{\pi_1,\dots,\pi_m\}$ of nonempty sequences
$\pi_p=(\pi_{p1},\dots,\pi_{p,n_p})$, 
such that the family
of sets $\{\pi_{p1},\dots,\pi_{p,n_p}\}$, $1\leq p\leq m$ 
forms a partition of $A$.
\end{definition}

For a fixed positive constant $\gamma_0$ let us consider a vector 
space $\widetilde{\Gammaqb}(\A)=\bigoplus_{n\geq 0} \A^{\stimes n}$ (where 
$\A^{\stimes n}$ denotes the symmetric tensor power) with a sesquilinear form
defined by 
\begin{equation} \label{eq:iloczyn} 
\langle \chi_1\otimes\cdots\otimes\chi_k,
\psi_1\otimes\cdots\otimes\psi_l\rangle= 
\end{equation}
$$=\delta_{kl} \frac{2^k}{k!}
\sum_{\{\pi_1,\dots,\pi_m\}} \prod_{1\leq p\leq m} 
\frac{\gamma_0}{n_p}  \mu(\chi_{\pi_{p1}}\gwia \psi_{\pi_{p1}}^{}\cdots
\chi_{\pi_{pn_p}}\gwia \psi_{\pi_{pn_p}}^{}), $$
where the sum is taken over all ordered partitions $\pi$ of the set $\{1,\dots,n\}$.

Please note that this sesquilinear form is well--defined on the full 
tensor power $\A^{\otimes n}$, however we shall usually use
it on the symmetric tensor power $\A^{\stimes n}$.

In the sum $\widetilde{\Gammaqb}=\bigoplus_{n\geq  0} \A^{\stimes n}$ 
appears a
summand $\A^{\stimes 0}$ which should be understood as a one-dimensional 
Hilbert
space $\C \Omega$ where $\Omega$ is a unital vector called vacuum.

The Hilbert space $\Gammaqb$, a completion of $\widetilde{\Gammaqb}$ will be 
called bosonic quadratic Fock space. In the following by 
$\A^{\stimes k}$ we shall mean the completion of the symmetric tensor power $\A^{\stimes k}$ with 
respect to the scalar product (\ref{eq:iloczyn}).

\begin{question}
For the sesqilinear form (\ref{eq:iloczyn}) all
algebraic considerations of this section 
hold even if the algebra $\A$ is not commutative. If this case we only
have to assume that the state $\mu$ is tracial and we 
have to replace the number operator (\ref{eq:definicjan}) by a pair of
left and right number operators. 
Unfortunately, in this general situation the form (\ref{eq:iloczyn})
is not always positively definite. Is it possible to find some nontrivial examples
of noncommutative 
algebras $\A$ with tracial states $\mu$ such that (\ref{eq:iloczyn}) is positively 
definite?
\end{question}

For $\psi\in\A$ we define the action of the quadratic creation, annihilation 
and number operators
on simple tensors by
\begin{equation} b\gwia_\psi(\chi_1\otimes\cdots\otimes\chi_k)= \sum_{0\leq 
i\leq k} 
\chi_1\otimes\cdots\otimes\chi_{i}\otimes\psi\otimes\chi_{i+1}\otimes 
\cdots\otimes\chi_k, \label{eq:definicjabgwia} \end{equation}
\begin{equation} b_\psi(\chi_1\otimes\cdots\otimes\chi_k)= 
2 \gamma_0\ \mu(\psi\gwia \chi_1)\
\chi_2\otimes\cdots\otimes\chi_k+ \end{equation}
$$+2\!\sum_{2\leq i \leq k} 
\chi_2\otimes\dots\otimes\chi_{i-1}\otimes( \chi_i \psi\gwia
\chi_{1})\otimes\chi_{i+1}\otimes\cdots\otimes\chi_k, $$ 
\begin{equation} n_\psi(\chi_1\otimes\cdots\otimes\chi_k)=\sum_{1\leq i\leq 
k} \chi_1\otimes\cdots\chi_{i-1}\otimes (\psi 
\chi_i)\otimes\chi_{i+1}\otimes\cdots\otimes\chi_k,
\label{eq:definicjan} \end{equation} 
for $k\geq 1$ and their action on the vacuum by
\begin{equation} b\gwia_\psi(\Omega)=\psi, \qquad
 b_\psi(\Omega)=0,\qquad n_\psi(\Omega)=0. 
\end{equation}
Please note that simple tensors are in 
general not elements of the symmetric tensor power of $\A$. However, by 
linearity these definitions extend to a dense subspace of the symmetric 
tensor power $\A^{\stimes n}$. What is important, the range of operators 
$b_\phi:\A^{\stimes n}\rightarrow \A^{\stimes(n-1)}$, 
$b_\phi\gwia:\A^{\stimes n}\rightarrow \A^{\stimes(n+1)}$, 
$n_\phi:\A^{\stimes n}\rightarrow \A^{\stimes n}$ is again a symmetric power 
of $\A$.

A difficulty arises from the fact that such defined operators are not bounded. 
For example we shall not claim that $b_\phi\gwia$ is an adjoint of $b_\phi$ 
because such a statement is not easy to prove since it demands careful 
discussion of domains of operators. It seems that in order to do this we 
would have to define these operators on some analogue of exponential domain 
of Hudson and Parthasarathy \cite{HuP} 
in a less intuitive way. Similarly commutation 
relations will hold only in a restricted sense.

\begin{theorem}
Operators $b_\phi$, $b\gwia_\phi$, $n_\phi$ fulfill the following 
operator norm estimates with respect to the scalar product 
(\ref{eq:iloczyn}):
\begin{equation} \label{eq:est1} \left\|b_\phi:\A^{\stimes 
k}\rightarrow\A^{\stimes (k-1)}\right\| \leq \sqrt{2k}\ \Big( 
\sqrt{\gamma_0}\ \|\phi\|_\Ldwa+(k-1) \|\phi\|_\Linf 
\Big), \end{equation} 
\begin{equation} \label{eq:est2} \left\|b\gwia_\phi:\A^{\stimes 
(k-1)}\rightarrow\A^{\stimes k}\right\| \leq \sqrt{2k}\ \Big( 
{\sqrt{\gamma_0}}\ \|\phi\|_\Ldwa+(k-1) \|\phi\|_\Linf 
\Big), \end{equation} 
\begin{equation} \label{eq:est3} 
\left\|n_\phi:\A^{\stimes k}\rightarrow\A^{\stimes k}\right\| 
\leq k\ 
\|\phi\|_\Linf. \end{equation} \end{theorem} 
\begin{proof}
Let us consider a map $\A^{\otimes k}\rightarrow\A^{\otimes (k-1)}$ 
defined on simple tensors by 
$\psi_1\otimes\cdots\otimes\psi_k\mapsto 2\gamma_0 \langle 
\phi,\psi_1\rangle\psi_2\otimes\cdots\otimes \psi_k$. It is easy to see 
that the operator norm of this map does not exceed $\sqrt{2k\gamma_0}\ 
\|\phi\|_\Ldwa$.

And now, for any $i$ let us consider a map $\A^{\otimes 
k}\rightarrow\A^{\otimes (k-1)}$ defined on simple tensors by 
$\psi_1\otimes\cdots\otimes\psi_k\mapsto
2\psi_2\otimes\cdots\otimes\psi_{i-1}\otimes(
\psi_i \phi\gwia \psi_1)\otimes\psi_{i+1}\otimes\cdots\otimes \psi_k$. 
 It is easy to see that the operator norm of this map does not exceed 
$\sqrt{2k}\ \|\phi\|_\Linf$.

The sum of these maps is equal to $b_\phi$, which shows the estimate 
(\ref{eq:est1}).

The estimation (\ref{eq:est2}) follows from (\ref{eq:est1}) because $b_\phi$ 
is an adjoint of $b\gwia_\phi$ what will be proven in Theorem 
\ref{theo:adjoint} and therefore their norms are equal.

The inequality (\ref{eq:est3}) is obvious.
\qed \end{proof}

This theorem allows us to define the action on $\A^{\stimes k}$ of operators 
$a_\phi$, $a_\phi\gwia$  for all $\phi\in\Ldwa(M,d\mu)\cap \Linf(M,d\mu)$ 
and of operator $n_\phi$ for all $\phi\in\Linf(M,d\mu)$.

\begin{theorem} \label{theo:adjoint} For any
$\zeta\in\Ldwa(M,d\mu)\cap\Linf(M,d\mu)$ operators $b_\zeta$ and $b\gwia_\zeta$ are 
adjoint in the sense that $$\langle b_\zeta \Psi, \Phi \rangle=\langle \Psi, 
b\gwia_\zeta \Phi \rangle $$ 
for all $\Psi\in\A^{\stimes k}$, $\Phi\in\A^{\stimes l}$.
For any $\phi\in\Linf(M,d\mu)$ the adjoint of $n_\phi$ is equal to 
$n_{\phi\gwia}$ 
in the sense that
$$\langle n_\zeta \Psi, \Phi \rangle=\langle \Psi, 
n_{\zeta\gwia} \Phi \rangle, $$ 
for all $\Psi\in\A^{\stimes k}$, $\Phi\in\A^{\stimes l}$.
\end{theorem}
\begin{proof}
Let us consider $\Psi=\sum_M
\psi^M_0\otimes\cdots\otimes\psi^M_{k-1}\in\A^{\stimes k}$ and $\Phi=\sum_N 
\phi^N_1\otimes\cdots\otimes\phi^N_l\in\A^{\stimes l}$.
Since $\Psi$ is a symmetric tensor the value of a scalar product
$\left\langle \Psi,\sum_N 
\phi^N_1\otimes\cdots\otimes\phi^N_{i-1}\otimes\zeta\otimes\phi^N_{i} 
\otimes\cdots\otimes\phi^N_l\right\rangle$
does not depend on $i$. This implies that
$$\langle\Psi,b\gwia(\zeta)\Phi\rangle=(l+1) 
\langle\Psi,\zeta\otimes\Phi\rangle.$$
We can split the sum in the definition (\ref{eq:iloczyn}) of 
$\langle\Psi,\zeta\otimes\Phi\rangle$ into two parts: over ordered 
partitions $\pi$ 
of the set $\{0,1,\dots,k-1\}$ which contain a block consisting of a single 
element $0$ and all the others ordered partitions. Since the state $\mu$ is 
tracial we have
$$\langle \Psi,\zeta\otimes\Phi\rangle
=\delta_{k,l+1} \frac{2^k}{k!} \sum_M \sum_N \Biggl[
 \gamma_0\ \mu(\psi^{M\star}_0
\zeta)\times$$ $$\times  \sum_{\{\pi_1,\dots,\pi_m\}}
 \prod_{1\leq p\leq m} \frac{\gamma_0}{n_p} 
\mu(\psi_{\pi_{p1}}^{M\star}\phi_{\pi_{p1}}^N 
\cdots\psi_{\pi_{p,n_p}}^{M\star} \phi_{\pi_{p,n_p}}^{N})+$$
$$+ \sum_{\{\pi_1,\dots,\pi_m\}} \sum_{1\leq q\leq m} \gamma_0\ 
\mu(\psi_0^M{}\gwia \zeta
\psi_{\pi_{q1}}^{M\star} \phi_{\pi_{q1}}^N \cdots
\psi_{\pi_{q,n_q}}^{M\star} \phi_{\pi_{q,n_q}}^{N}) 
\times $$ $$\times 
\prod_{1\leq p\leq m, p\neq q} \frac{\gamma_0}{n_p} 
\mu(\psi_{\pi_{p1}}^{M\star}\cdots\psi_{\pi_{p,n_p}}^{M\star} 
\phi_{\pi_{p1}}^{N}\cdots\phi_{\pi_{p,n_p}}^{N}) \Biggr],$$ where the sums over 
$\pi$ are taken over all ordered partitions $\pi$ of the set $\{1,\dots,k-1\}$. 

Note that for any nonempty subset $A$ of the set $\{1,\dots,k-1\}$ we have
$$ \sum_{\pi_q} 
\mu(\psi_0^M{}\gwia \zeta
\psi_{\pi_{q1}}^{M\star} \phi_{\pi_{q1}}^N \cdots
\psi_{\pi_{q,n_q}}^{M\star} \phi_{\pi_{q,n_q}}^{N}) 
=\sum_{\pi_q} \sum_{1\leq r\leq n_q} \frac{1}{n_q}\times $$ $$
\times \mu(\psi_{\pi_{q1}}^{M\star} \phi_{\pi_{q1}}^N
\cdots\psi_{\pi_{q,r-1}}^{M\star} \phi_{\pi_{q,r-1}}^N
(\psi^M_{\pi_{q,r}} \zeta\gwia \psi^M_0)\gwia \phi^N_{\pi_{q,r}} 
\psi_{\pi_{q,r+1}}^{M\star} \phi_{\pi_{q,r+1}}^N\cdots
\psi_{\pi_{q,n_q}}^{M\star} \phi_{\pi_{q,n_q}}^N), $$
where the sums are taken over all sequences $\pi_q=(\pi_{q,1},\dots,
\pi_{q,n_q})$ such that each of the elements of $A$ appears in $\pi_q$
exactly once.

Now, 
it is easy to see that $$\langle \Psi,b\gwia_\zeta\Phi\rangle
=2 \delta_{k,l+1} \left[ \gamma_0 \sum_M \langle \psi^M_0,\zeta\rangle 
\langle \psi^M_1\otimes\cdots\otimes\psi^M_k,\Phi\rangle+ \right. $$
$$\left. +\sum_i \sum_M \langle 
\psi^M_1\otimes\cdots\otimes\psi^M_{i-1}\otimes (\psi^M_i 
\zeta\gwia \psi^M_0)\otimes \psi^M_{i+1}\otimes\cdots\psi^M_k,
\Phi\rangle \right],$$
which proves the first part of the theorem.

The proof of the fact that the adjoint of $n_\phi$ is equal to
$n_{\phi\gwia}$ is very simple and we shall omit it. 
\qed \end{proof}

\begin{theorem} \label{thoe:komutacja}
For any $\phi,\psi\in\Ldwa(\A)\cap\Linf(\A)$, $\zeta,\eta\in\Linf(\A)$ and 
$\Phi\in\A^{\stimes k}$ we have 
\begin{equation} \label{eq:kom1} [b\gwia_\phi, b\gwia_\psi] \Phi=0, \qquad 
[b_\phi, b_\psi] \Phi=0, \end{equation}
\begin{equation} [b_\phi,b\gwia_\psi] \Phi=(2 \gamma_0 \langle 
\phi,\psi\rangle +4 n_{\phi\gwia \psi}) \Phi, \end{equation}
\begin{equation}[n_\zeta,b\gwia_\psi]\Phi=2b\gwia_{\zeta \psi} \Phi, \qquad
 [b_\psi,n_\zeta]\Phi=2b_{\zeta\gwia \psi} \Phi.
\end{equation}
\end{theorem}
\begin{proof} 
Since the definitions of creation quadratic operators and standard creation
operators coincide, two quadratic creation operators commute.
Qua\-dra\-tic annihilation operators are their adjoints so they commute with
each other as well.

Let us consider two auxiliary annihilation operators
$$\hat{b}_\psi(\chi_1\otimes\cdots\otimes\chi_k)= 2 \gamma_0\ \mu(\psi\gwia \chi_1)\
\chi_2\otimes\cdots\otimes\chi_k,$$
$$\tilde{b}_\psi(\chi_1\otimes\cdots\otimes\chi_k)= 
2\!\sum_{2\leq i \leq k} 
\chi_2\otimes\dots\otimes\chi_{i-1}\otimes( \chi_i \psi\gwia
\chi_{1})\otimes\chi_{i+1}\otimes\cdots\otimes\chi_k. $$ 
We have $b_\psi=\hat{b}_\psi+\tilde{b}_\psi$.

The definition of $\hat{b}$ coincides up to a factor with the definition of
the standard annihilation operator, therefore
$$[\hat{b}_\phi,b\gwia_\psi]=2 \gamma_0 \langle \phi,\psi\rangle.$$

It is easy to see that there are exactly two terms in the commutator which
do not cancel:
$$[\tilde{b}_\phi,b\gwia_\psi](\chi_1\otimes\cdots\otimes\chi_k)=
2\gamma_0 (\psi\phi\gwia\chi_1+\chi_1\phi\gwia\psi)\otimes\chi_2\otimes\cdots\otimes\chi_k,$$
which is equal to the action of $4 \gamma_0 n_{\psi\phi\gwia}$. 
If we do not assume
that $\A$ is commutative we have to replace $n$ by an appropriate 
sum of left and right
multiplication operators.\qed \end{proof}

\section{Another Representation of the Quadratic Bosonic Fock Space}
\label{sec:another}
The construction from the previous subsection 
can be presented in a more direct way. 
Let us consider an isomorphism
\newcommand{\alg}{_{\rm{alg}}}
$C(M)\otimes\cdots\otimes C(M)=C\alg (M\times\cdots\times M)$, where 
$C\alg(M^n)$ denotes 
the space of continuous functions on 
$M^n=M\times\cdots\times M$ which are finite sums 
of simple tensors. The multiplication map $\A^{\otimes n}\ni 
x_1\otimes\cdots \otimes x_n \mapsto 
x_1 \cdots x_n\in\A$ under this isomorphism is equal to the diagonal 
map $C\alg (M^n)\ni f\mapsto \Delta f  \in C(M)$, 
where $(\Delta f)(x)=f(x,x,\dots,x)$ for any $x\in M$. 

For any ordered 
partition $\pi=\{\pi_1,\dots,\pi_k\}$ of the set $\{1,\dots,n\}$ let 
$\Delta_\pi:M^k \rightarrow M^n$ be an embedding of $M^k$ onto the diagonal 
of $M^n$ defined by partition $\pi$: 
$$\Delta_\pi(x_1,\dots,x_k)=(y_1,\dots,y_n), $$ where $y_{r}=x_s$ for 
$r\in\pi_s$. $\Delta_{\pi}\gwia(\mu^{\otimes m})$ denotes 
the pull--back of the 
measure $\mu^{\otimes m}$ on $M^m$ onto a multidiagonal of $M^k$ defined by 
$\Delta_\pi$, namely
$$\int_{M^k} f(x_1,\dots,x_k)\ d\Delta_{\pi}\gwia(\mu^{\otimes m})=
\int_{M^m} f[\Delta_\pi(y_1,\dots,y_m)] d\mu(y_1)\cdots d\mu(y_m).$$
Note that however the function $\Delta_\pi$ depends on the choice of order
of blocks of partition $\pi$, the pull--back measure
$\Delta_{\pi}\gwia(\mu^{\otimes m})$ does not depend on it.

Therefore the 
scalar product (\ref{eq:iloczyn}) can be represented as
$$ \langle \Phi,\Psi \rangle= 
\delta_{kl} \int_{M^k} \overline{\Phi(x_1,\dots,x_k)} 
\Psi(x_1,\dots,x_k)\ d\mu_k(x_1,\dots,x_k)$$
for $\Phi\in C\alg(M^k)$, $\Psi\in C\alg(M^l)$, where the measure $\mu_k$ on 
$M^k$ is given by
$$\mu_k=\frac{2^k}{k!} \sum_{\{\pi_1,\dots,\pi_m\}} 
\frac{\gamma_0^m}{|\pi_1|\cdots|\pi_m|}
\Delta_{\pi}\gwia(\mu^{\otimes m}).$$ 
%
In other words: the measure $\mu_k$ on $M^k$ is a sum of the product measure 
on $M^k$ and of product measures with supports on all multidiagonals of 
$M^k$.

The operators defined in the last section in this context are represented 
as follows: 
\begin{equation} (b\gwia_\phi \Psi)(x_1,\dots,x_{n+1})=\sum_i \phi(x_i) 
\Psi(x_1,\dots,x_{i-1},x_{i+1},\dots,x_{n+1}), \end{equation}
\begin{equation} (b_\phi 
\Psi)(x_1,\dots,x_n)=2 \gamma_0 \int_M \overline{\phi(x_{n+1})} 
\Psi(x_1,\dots,x_{n+1}) d\mu(x_{n+1})+\end{equation} $$+2 \sum_i 
\overline{\phi(x_i)} 
\Psi(x_1,x_2,\dots,x_{i-1},x_{i},x_{i},x_{i+1},\dots,x_n),$$
\begin{equation}(n_\phi \Psi)(x_1,\dots,x_n)= \Psi(x_1,\dots,x_n) \sum_i 
\phi(x_i).\end{equation}

\section{Quadratic and Linear Bosonic White Noise}
\label{sec:quadraticandlinear}
It is natural to ask if it is possible to incorporate both quadratic white 
noise operators $b_\phi$, $b\gwia_\phi$, $n_\phi$ and linear white noise 
operators $a_\phi$, $a\gwia_\phi$ to the same algebra. We postulate 
relations (\ref{eq:t0prim})--(\ref{eq:t2prim}) of quadratic white noise, a 
relation of white noise
$$ [a_\phi,a\gwia_\psi]=\langle \phi,\psi\rangle, $$
and some relations linking quadratic and linear noises, among which we shall 
mention only 
$$ [a_\phi,b\gwia_\psi]=2 a\gwia_{\phi\gwia \psi}, \qquad
 [b_\phi,a\gwia_\psi]=2 a_{\phi \psi\gwia}.  $$
We shall prove now that in general it is impossible to find a Fock 
representation of these 
relations.

Let $X$ be a measurable subset of $M$. Let $0<\mu(X)=l<\infty$ and let 
$\chi(x)=1$ for $x\in X$ and $\chi(x)=0$ otherwise. By rewriting operators 
in the normal order we have for any $c\in\R$,
 $$\langle (c a\gwia_\chi a\gwia_\chi+b\gwia_\chi)\Omega, (c 
a\gwia_\chi a\gwia_\chi+b\gwia_\chi)\Omega\rangle=
\langle \Omega,(c a_\chi a_\chi+b_\chi)(c a\gwia_\chi 
a\gwia_\chi+b\gwia_\chi)\Omega\rangle=$$ $$=
2c^2 \langle\chi,\chi\rangle^2+2 \gamma_0 \langle\chi,\chi\rangle+
2c \langle \chi^2,\chi\rangle+2c \langle \chi,\chi^2\rangle=
2 c^2 l^2+4c l+2 \gamma_0 l.$$ 
It is easy to see that for $l<\frac{1}{\gamma_0}$ this expression takes 
negative values.

This is can be interpreted as another manifestation of the constant
$\frac{1}{\gamma_0}$ which describes the lengthscale, under which a 
quadratic white noise loses its physical meaning.

\section{Free Quadratic White Noise} \label{sec:free}
\subsection{Free commutation relations}
For the free case $q=0$ the coefficient standing at number operator in 
Eq.\ (\ref{eq:sss}) is equal to $0$ so this equation is equivalent to 
the commutation relations of free creation and annihilation operators. 
However if we redefine annihilation and creation operators
by mulitiplying them by $\frac{1}{\sqrt{q}}$
and take the limit $q\rightarrow 0$ and $ql=\frac{1}{\gamma}$ 
we obtain
\begin{equation} b_\phi b\gwia_\psi=\gamma \langle \phi,\psi \rangle 
+n_{\bar{\phi} \psi}. \label{eq:tbis} \end{equation}

The simple calculation for $q=0$,
$$(a_{\frac{1}{\sqrt{l}} \chi_i}\gwia a_{\frac{1}{\sqrt{l}} \chi_i}) 
(a\gwia_{\frac{1}{\sqrt{l}} \chi_j} )^2=\delta_{ij} 
(a\gwia_{\frac{1}{\sqrt{l}} \chi_j} )^2,$$ 
$$(a_{\frac{1}{\sqrt{l}} \chi_i}\gwia a_{\frac{1}{\sqrt{l}} \chi_i}) 
(a_{\frac{1}{\sqrt{l}} \chi_j}\gwia a_{\frac{1}{\sqrt{l}} \chi_j})
 =\delta_{ij} 
 (a_{\frac{1}{\sqrt{l}} \chi_i}\gwia a_{\frac{1}{\sqrt{l}} \chi_i})$$ 
motivates us to postulate the other free commutation relations 
\begin{equation} n_\phi 
b\gwia_\psi=b\gwia_{\phi\psi},\qquad b_\psi n_\phi=b_{\psi 
\phi\gwia},\qquad n_\phi n_\psi=n_{\phi \psi}. \end{equation}

Therefore, heuristically quadratic free white noise defined like this 
could be interpreted as a square of free white noise with small violations 
of freeness in the limit $q\rightarrow 0$.
However, since we take the limit $l\rightarrow\infty$ it is impossible to repeat 
the arguments from Subsect. \ref{subsec:variation} and it should be
stressed that this interpretation is very informal.

\subsection{Realization of quadratic free white noise}
Let $\A$ be an associative $\star$-algebra and $\mu:\A\rightarrow\C$ be a 
state. 
Let us consider a pre--Hilbert 
space $\widetilde{\Gammaqf}(\A)=\bigoplus_{n\geq 0} \A^{\otimes n}$ with a 
scalar product defined by
\begin{equation} \langle 
\psi_1\otimes\cdots\otimes\psi_l, \chi_1\otimes\cdots\otimes\chi_k\rangle= 
\label{eq:wolny} \end{equation}
 $$=\delta_{kl}
\sum_{m\geq 1} \sum_{(n_0,\dots,n_m)} \prod_{1\leq p\leq m} \gamma
\mu(\psi_{n_{p}}\gwia \psi_{n_p-1}\gwia\cdots 
\psi_{n_{p-1}+1}\gwia
\chi_{n_{p-1}+1} \chi_{n_{p-1}+2} \cdots \chi_{n_{p}}),$$ 
where 
the sum is taken over all increasing sequences of natural numbers 
$(n_0,n_1,\dots,n_m)$ such that $n_0=0$ and $n_m=k$ what corresponds to all 
Boolean partitions of a set $\{1,\dots,k\}$, i.e.\ partitions 
into blocks of consecutive elements
$\{ \{1,2,\dots,n_1\}, \{n_1+1,n_1+2,\dots,n_2\},\dots, 
\{n_{m-1}+1,n_{m-1}+2,\dots,n_m\}\}$. 

The completion of $\widetilde{\Gammaqf}(\A)$ will be called the 
quadratic free Fock
space and will be denoted by ${\Gammaqf}(\A)$.

 For $\psi\in\A$ we define the action of the 
quadratic creation operators on $\widetilde{\Gammaqf}$ by
$$b\gwia_\psi(\chi_1\otimes\cdots\otimes\chi_k)= 
\psi\otimes\chi_1\otimes\cdots\otimes\chi_k,$$
$$b_\psi(\chi_1\otimes\cdots\otimes\chi_k)= 
\gamma \mu(\psi\gwia \chi_1)\
\chi_2\otimes\cdots\otimes\chi_k+(\psi\gwia \chi_1 \chi_2)
\otimes\chi_{3}\otimes \cdots\otimes\chi_k, $$
$$n_\psi=(\psi \chi_1)\otimes\chi_2\otimes\cdots\otimes\chi_k.$$

On the algebra $\A$ we shall introduce (noncommutative) $\Ldwa$ and $\Linf$ 
norms by
$$\|x\|_\Ldwa=\sqrt{\mu(x\gwia x)},$$
$$\|x\|_\Linf=\sup_{y\in\A,\ \|y\|_{\Ldwa}=1} |\mu(xy)|.$$
 \begin{theorem} For quadratic free operators the following estimations 
hold: $$\left\|b\gwia_\phi\right\|=\left\|b_\phi\right\|\leq \sqrt{\gamma}
\|\phi\|_{\Ldwa}+\|\phi\|_\Linf, $$ 
$$\|n_\phi\|\leq \|\phi\|_\Linf.$$
\end{theorem}

\begin{theorem} The operators $b_\phi$ and $b_\phi\gwia$ are adjoint. The 
adjoint to $n_\phi$ is equal to $n_{\phi\gwia}$.
 \end{theorem}
\begin{proof}
First note that the summands in 
(\ref{eq:wolny}) can be split into two groups: those for which 
the first component of 
the partition defined by $(n_i)$ is and those for which 
is not a single element. 
Therefore for $\Psi=\psi_1\otimes\cdots\otimes\psi_{k}$, 
$X=\chi_0\otimes\cdots\otimes \chi_{k}$ we have
 $$\langle b\gwia_{\phi} \Psi, X\rangle=\langle \phi\otimes 
\psi_1\otimes\cdots\otimes\psi_{k},  \chi_0\otimes\cdots\otimes 
\chi_{k}\rangle=\gamma \mu(\phi\gwia \chi_0)\times $$ $$\times \sum_{m\geq 1} 
\sum_{(n_0,\dots,n_m)} \prod_{1\leq p\leq m} 
\gamma \mu(\psi_{n_{p}}\gwia \psi_{n_{p}-1}\gwia \cdots \psi_{n_{p-1}+1}\gwia 
\chi_{n_{p-1}+1} \chi_{n_{p-1}+2} 
\cdots \chi_{n_{p}})+$$
$$+\sum_{m\geq 1} \gamma^{m} 
\sum_{(n_0,\dots,n_m)} \mu[\psi_{n_1}\gwia \psi_{n_1-1}\gwia \cdots 
\psi_1\gwia \phi\gwia \chi_{0} \cdots \chi_{n_1} ]  \times $$ $$\times 
\prod_{2\leq p\leq m} \mu[\psi_{n_{p}}\gwia 
\psi_{n_{p}-1}\gwia\cdots\psi_{n_{p-1}+1}\gwia   \chi_{n_{p-1}+1} \chi_{n_{p-1}+2} 
\cdots \chi_{n_{p}} ]= $$
$$=\langle \psi_1\otimes\cdots\otimes\psi_k, \gamma \mu(\phi\gwia \chi_0) 
 \chi_1\otimes\dots\otimes \chi_k \rangle + $$ 
$$+\langle\psi_1\otimes\cdots\otimes\psi_k, (\phi\gwia \chi_0 \chi_1)
\otimes\chi_{2}\otimes \cdots\otimes\chi_k\rangle, $$
where 
the sums are taken over all increasing sequences of natural numbers 
$(n_0,\dots,n_m)$ such that $n_0=0$ and $n_m=k$.

The proof of the fact that $n_\phi$ is adjoint to $n\gwia_\phi$ is 
straightforward and we shall omit it. 
\qed \end{proof}

\begin{theorem}
The following operator equalities hold for all $\phi,\psi\in\Ldwa\cap\Linf$ 
and $\zeta, \eta\in\Linf$:
\begin{equation} b_\psi b\gwia_\phi=\gamma \mu(\psi\gwia \phi) + n_{\psi\gwia 
\phi}, \label{eq:free1} \end{equation} 
\begin{equation} n_\zeta b_\phi\gwia=b_{\zeta \phi}\gwia,
\qquad b_\psi n_\zeta=b_{\zeta\gwia \psi}, \end{equation}
\begin{equation} n_\zeta n_\eta=n_{\zeta \eta}. 
\label{eq:free3} \end{equation} \end{theorem}

\subsection{Free quadratic Fock space and free probability}
In this subsection we shall present some properties of the free quadratic Fock space 
related to the free probability of Voiculescu \cite{V}. 
\newcommand{\F}{{\mathcal F}} \begin{definition}
%
A noncrossing partition is a partition $\pi=\{\pi_1,\dots,\pi_k\}$ of a set 
$\{1,\dots,n\}$ such that there do not exist numbers $1\leq a<b<c<d\leq n$ 
such that $a,c\in\pi_r$, $b,d\in\pi_s$ and $r\neq s$. 
\end{definition}
\begin{theorem} 
Let $\A$ be an associative $\star$--algebra with a state $\mu$.
 For $Q_s(\phi)=b\gwia_\phi+b_{\phi\gwia}+s n_\phi$ we have 
$$\tau[Q_s(\phi_1) \cdots Q_s(\phi_{k})]=$$ 
$$=\sum_{\pi=\{\pi_1,\dots,\pi_k\}} \prod_{1\leq i\leq k} 
\gamma  \mu(\phi_{\pi_{i1}} \cdots \phi_{\pi_{i,n_i}}) \sum_{1\leq 
l\leq \frac{n_i-2}{2}} \frac{1}{l+1} {{2l}\choose{l}} {n_i-2 \choose 2l} 
s^{n_i-2l-2},$$
{\sloppy
where the sum is taken over all 
noncrossing partitions $\{\pi_1,\dots,\pi_k\}$; 
$\pi_i=\{\pi_{i,1},\dots,\pi_{i,n_i} \}$, $\pi_{i,1}<\cdots<\pi_{i,n_i}$. }

The free cumulants \cite{KS} are therefore 
$$k_n(\phi_1,\dots,\phi_n)=
\gamma  \mu(\phi_{1} \cdots \phi_{n}) \sum_{1\leq 
l\leq \frac{n-2}{2}} \frac{1}{l+1} {{2l}\choose{l}} {n-2 \choose 2l} 
s^{n-2l-2}.$$
\end{theorem}
\begin{proof} 
\newcommand{\btrzy}{{\tilde{b}}}
\newcommand{\bjed}{{\hat{b}}}
Let us consider two auxiliary annihilation operators
$$\bjed_\phi (\psi_1\otimes\cdots\otimes\psi_k)=\gamma \mu(\phi\gwia \psi_1)\ \psi_2
\otimes\cdots\otimes\psi_k,$$
$$\btrzy_\phi (\psi_1\otimes\cdots\otimes\psi_k)=(\phi\gwia \psi_1 \psi_2) 
\otimes \psi_3 \otimes \cdots \otimes \psi_k.$$
Therefore $Q_s(\phi)=b\gwia_\phi+s n_\phi+\btrzy_{\phi\gwia}+\bjed_{\phi\gwia}$ and
$\tau[Q_s(\phi_1)\cdots Q_s(\phi_k)]$ is a sum of $4^k$ summands
each equal to the state $\tau$ acting on a product of operators $b\gwia$, 
$sn$, $\btrzy$ and $\bjed$. Each of these summands is of the form
 $\prod_{1\leq i\leq k} \gamma \mu(\phi_{\pi_{i1}} \cdots 
\phi_{\pi_{i,n_i}})$ times a power of $s$. Furthermore, only expressions 
coming from noncrossing partitions can appear. Our question is: with which 
coefficient such a term comes in the $\tau[Q_s(\phi_1)\cdots Q_s(\phi_k)]$. 
We shall discuss the $4^{n_i}$ ways of choosing one of four operators 
$b\gwia$, $sn$, $\btrzy$ and $\bjed$ to be associated with each of the 
vectors $\phi_{\pi_{i1}},\dots,\phi_{\pi_{i,n_i}}$ forming a block of the 
partition $\pi$.

First note that there must be $n_i\geq 2$, and with the vector 
$\phi_{\pi_{i1}}$ must be associated the annihilator $\bjed$ and 
with the vector $\phi_{\pi_{i,n_i}}$ must be associated the creator 
$b\gwia$---otherwise such a summand does not contribute in the sum. There are 
remaining $n_i-2$ places on which we have to choose operators $b\gwia$, 
$\btrzy$ and $sn$. The number of 
creation operators $l$ on these places must be equal to the number of 
annihilators, the other $n_i-2-2l$ places must be occupied by number 
operators. There are $n_i-2 \choose 2l$ possibilites of choosing places on 
which number operators should be placed. The number of ways of choosing $l$ 
places among $2l$ places on which creation operators should act is
equal to the $l$ Catalan number $\frac{1}{n+1} {2l \choose l}$ \cite{GKP} 
which ends the proof. 
\qed \end{proof}
As a simple corollary we have the following 
\begin{theorem} Let $\A$ 
be an associative $\star$--algebra with a tracial state $\mu$ and let $s\in\R$.
Let $\F_s(\A)$ be a $\star$--algebra generated by operators $Q_s(\phi)$ 
for $\phi\in\A$ and 
by the identity operator. Then a state $\rho$ on $\F_s(\A)$ given by 
$\rho(X)=\langle \Omega,X\Omega\rangle$ is tracial. \end{theorem}

\begin{theorem} Let $\A$ be an associative $\star$--algebra with a tracial 
state $\mu$, $s\in\R$ and let $(\A_i)$ be a family of $\star$--subalgebras of $\A$ 
such that $xy=0$ for any $x\in\A_i$, $y\in\A_j$ and $i\neq j$. Then 
subalgebras $\F_s(\A_i)$ of the algebra $\F_s(\A)$ are free with 
respect to the state $\rho$.
\end{theorem}
\begin{proof}
By the definition of freeness we have to prove that
for any sequence $i_1,\dots,i_n$ of indexes such that 
the consecutive indexes are not equal $i_k\neq i_{k+1}$ for $1\leq k\leq 
n-1$ and a sequence $(X_k)$ such that $X_k\in\F_s(A_{i_k})$ and $\rho(X_k)=0$ 
for $1\leq k\leq n$ we have $\rho(X_1\cdots X_n)=0$.

Each of the operators $X_k$ can be written as a sum of normally ordered
operators, i.e.\ as a sum of products of the type $b\gwia_{\phi_1} \cdots 
b\gwia_{\phi_p} b_{\psi_1} \cdots b_{\psi_q}$ or $b\gwia_{\phi_1} \cdots 
b\gwia_{\phi_p} n_\zeta b_{\psi_1} \cdots b_{\psi_q}$. 
The assumption 
$\rho(X_k)=0$ implies that in this sum the multiplicity of the identity 
operator does not appear. By distributing we see that the
product $X_1\cdots X_n$ can be 
written as a sum of many summands, each being a product of normally ordered 
products.

The commutation relations (\ref{eq:free1})--(\ref{eq:free3}) show that for 
each of 
these summands if one is not normally ordered then it is equal to zero. On the 
other hand, a vacuum expectation of any normally ordered product (containing 
no multiplicity of identity) is equal to $0$.
\qed \end{proof}


%
%
%
%
%
%
%
%
%
%
%
%

\section{Acknowledgements}
The author would like to thank Prof.\ L.\ Accardi and M.\ Skeide for a
very inspiring presentation and
Centro Internazionale per la Ricerca Mathematica, Consiglio 
Nazionale della Ricerche and University of Roma Tor Vergata for giving 
opportunity to participate in the Volterra International School {\it White 
Noise Approach to Quantum Stochastic Calculi and Quantum Probability} in 
Trento, July 1999.

\end{document}